# CCAT-prime: a novel telescope for submillimeter astronomy.

Stephen C. Parshley[*a], Jörg Kronshage[b], James Blair[a], Terry Herter[a], Mike Nolta[c], Gordon J. Stacey[a], Andrew Bazarko[d], Frank Bertoldi[f], Ricardo Bustos[g], Donald B. Campbell[a], Scott Chapman[h], Nicholas Cothard[i], Mark Devlin[e], Jens Erler[f], Michel Fich[j], Patricio A. Gallardo[k], Riccardo Giovanelli[a], Urs Graf[l], Scott Gramke[b], Martha P. Haynes[a], Richard Hills[m], Michele Limon[e], Jeffrey G. Mangum[n], Jeff McMahon[o], Michael D. Niemack[k], Thomas Nikola[p], Markus Omlor[b], Dominik A. Riechers[a], Karl Steeger[b], Jürgen Stutzki[l], Eve M. Vavagiakis[k]

[a]Department of Astronomy, Cornell University, Ithaca, NY 14853, USA
[b]Vertex Antennentechnik, GmbH, 47198 Duisburg, Germany
[c]Canadian Institute for Theoretical Astrophysics, University of Toronto, Toronto, ON M5S 3H8, Canada
[d]Department of Physics, Princeton University, Princeton, NJ 08544, USA
[e]Department of Physics & Astronomy, University of Pennsylvania, Philadelphia, PA 19104, USA
[f]Argelander-Institute für Astronomie, Rheinische Friedrich-Wilhelms Universität Bonn, 53121 Bonn, Germany
[g]Universidad Católica de la Santísima Concepción, Alonso de Ribera 2850, Concepción, Chile
[h]Department of Physics and Atmospheric Science, Dalhousie University, Halifax, NS B3H 4R2, Canada
[i]Department of Applied and Engineering Physics, Cornell University, Ithaca, NY 14853, USA
[j]University of Waterloo, Waterloo, ON N2L 3G1, Canada
[k]Department of Physics, Cornell University, Ithaca, NY 14853, USA
[l]I. Physikalisches Institut, Universität zu Köln, 50937 Köln, Germany
[m]Cavendish Laboratory, University of Cambridge, Cambridge CB3 0HE, UK
[n]National Radio Astronomy Observatory, Charlottesville, VA 22903, USA
[o]Department of Physics, University of Michigan, Ann Arbor, MI 48109, USA
[p]Center for Astrophysics and Planetary Sciences, Cornell University, Ithaca, NY 14853, USA

## ABSTRACT

The CCAT-prime telescope is a 6-meter aperture, crossed-Dragone telescope, designed for millimeter and sub-millimeter wavelength observations. It will be located at an altitude of 5600 meters, just below the summit of Cerro Chajnantor in the high Atacama region of Chile. The telescope's unobscured optics deliver a field of view of almost 8 degrees over a large, flat focal plane, enabling it to accommodate current and future instrumentation fielding >100k diffraction-limited beams for wavelengths less than a millimeter. The mount is a novel design with the aluminum-tiled mirrors nested inside the telescope structure. The elevation housing has an integrated shutter that can enclose the mirrors, protecting them from inclement weather. The telescope is designed to co-host multiple instruments over its nominal 15 year lifetime. It will be operated remotely, requiring minimum maintenance and on-site activities due to the harsh working conditions on the mountain. The design utilizes nickel-iron alloy (Invar) and carbon-fiber-reinforced polymer (CFRP) materials in the mirror support structure, achieving a relatively temperature-insensitive mount. We discuss requirements, specifications, critical design elements, and the expected performance of the CCAT-prime telescope. The telescope is being built by CCAT Observatory, Inc., a corporation formed by an international partnership of universities. More information about CCAT and the CCAT-prime telescope can be found at www.ccatobservatory.org.

[*] scp8@cornell.edu; phone 1 (607) 255-4806; www.ccatobservatory.org

# 1 INTRODUCTION

Current state-of-the-art radio and submillimeter (submm) interferometers have amazing resolution and sensitivity. However, mapping large areas of the sky with such facilities is costly in terms of resources and time, since the mapping speed is limited by telescope throughput and detector count. Other specially designed survey telescopes can complement the exquisite precision of the large facilities (e.g. the Atacama Large Millimeter Array, ALMA) by mapping large regions quickly and identifying promising areas on the sky for detailed follow up work, as well as conducting valuable science programs in their own right. CCAT-prime is an exceptional survey telescope, designed to operate in the millimeter to submillimeter (100 GHz – 1.5 THz) range. With its 6-meter aperture, 7.8° field of view (FoV) at 3 mm wavelength, and superb site, it will enable studies that will explore the epoch of reionization, dark energy, inflation, as well as observations of diffuse atomic gas, large-scale filaments, giant molecular clouds and star-forming regions. Since submm radiation is absorbed by water vapor in Earth's atmosphere, the telescope must be located at a high, dry site. The observatory will be sited on Cerro Chajnantor in Chile, at 5600 meters above sea level. Additional descriptions of the overall science program and observatory can be found in Stacey et al. 2018[1].

Identifying a scientific niche with great potential and an optical design[2] that could deliver 10x more detectors on-sky than current observatories, we were motivated to design a telescope that could deliver the high performance needed to realize the scientific goals. The requirements for the telescope include a one half wavefront error (HWFE) of less than 11 μm, a pointing error (PE) of less than 1.4 arcsec, and an emissivity of less than 2.8% for wavelengths longer than 850 μm (350 GHz). For good scanning observing efficiency, the turnaround time must be < 2.5 sec for a maximum speed of 3 deg/sec in azimuth. Operations must be fully remote and regular maintenance is restricted to once per week. The telescope must provide significant volume and structural support for mounting science equipment in order to accommodate the multiple instruments needed to cover the various science cases. The adverse environmental conditions of the CCAT high site pose additional design challenges to developing a stable, robust telescope with a lifetime of at least 15 years.

The Simons Observatory[3] (SO) has adopted the design of the CCAT-prime telescope for their Large Aperture Telescope (LAT). The designs and specifications of the telescopes are nearly identical with the exception of the tighter tolerances required by CCAT-prime to enable submm observations (Table 1). The LAT will be located on a lower site, on the side of Cerro Toco, and will be dedicated to studying the cosmic microwave background in the range of 1-15 mm. The two projects are currently working together in the detailed design phase, and the telescopes will be constructed in parallel. CCAT-prime will undergo additional mirror alignment procedures at the factory while the LAT is shipped to Chile. Final assembly and site testing will be sequenced between the two telescope projects. First light for both telescopes is anticipated in 2021.

Table 1. Requirements for the CCAT-prime and SO LAT telescopes. All pointing requirements are in arcsec.

| Performance parameter | CCAT-prime requirement | CCAT-prime goal | SO LAT requirement |
|---|---|---|---|
| one half wavefront error | < 10.7 μm rms | < 7.1 μm rms | < 35 μm rms |
| blockage[1] | ≤ 1.0% | 0% | ≤ 1.0%[2] |
| turnaround time | < 2.5 sec | < 1.5 sec | - |
| blind pointing | ≤ 6.9 | ≤ 4.6 | - |
| offset pointing | ≤ 2.7 | ≤ 1.8 | ≤ 15 |
| scan pointing (sector 1, sector 2)[3] | ≤ 1.4, ≤ 1.9 | ≤ 0.9 | ≤ 4.0 |
| scan following error | ≤ 6.9 | ≤ 4.6 | ≤ 10.0 |
| pointing stability | ≤ 1.4 | ≤ 0.9 | ≤ 10.0 |

[1] Blockage is due to panel gaps since the optics are unobstructed.
[2] The SO LAT also has a goal of 0% blockage.
[3] Sector 1 and 2 refer to scan areas of 30° and 60° azimuth by 10° elevation, respectively.

This paper discusses requirements, specifications, critical design elements, and the expected performance of the CCAT-prime telescope. Simulated overall performance of HWFE, pointing, emissivity, and dynamics are presented. We also highlight a few major components of the system, including the preliminary reflector panel, mirror backup structure (BUS), elevation housing, and yoke. Additional details can be found in other manuscripts presented at this conference. For details on the optical design see Parshley et al. 2018[4]. A first light camera design is discussed in Vavagiakis et al. 2018[5].

## 2 TELESCOPE OVERVIEW

Figure 1 shows the mechanical design of the CCAT-prime telescope. It is roughly similar in appearance to a fork-structure (which we designate the yoke) alt-az mount without a reflector. Instead, the offset optics are contained inside what would normally be the secondary focus cabin. This cabin, referred to as the elevation housing, shelters the reflecting surfaces from wind and solar illumination. A shutter can be deployed to close the opening during inclement weather. The overall length, width, and height of the structure is approximately 23 x 8 x 16 meters, and the elevation axis is roughly 11 meters above the ground. The entire structure, including science equipment, weighs over 220 metric tons with approximately 200 tons moving in azimuth and 60 tons moving in elevation.

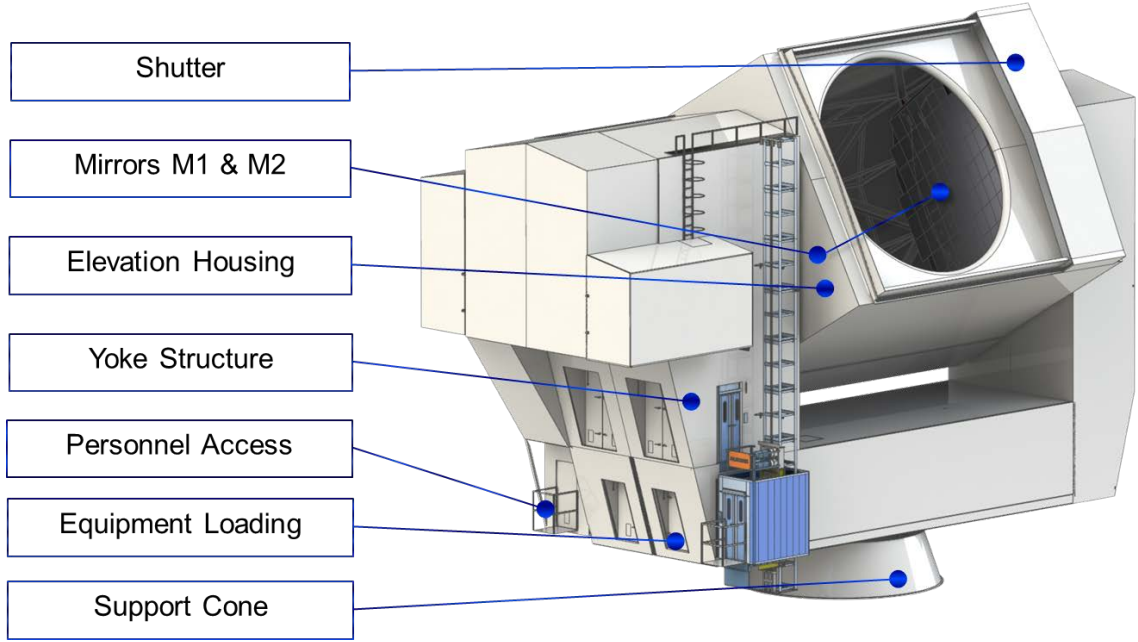

Figure 1. CCAT-prime telescope mechanical design model with major elements labeled. For reference, the opening in the elevation housing is approximately 7 m in diameter. The blue elevator on the side also gives a sense of scale.

### 2.1 Environment

The CCAT site is located just off of the peak of Cerro Chajnantor in the Atacama Desert in northern Chile, inside the Parque Astronómico de Atacama[6]. The general area is an excellent site for mm/submm astronomy. ALMA and a number of other observatories routinely operate on the Chajnantor plateau. The CCAT site is roughly 500 m above the plateau and shows a significant improvement in atmospheric transparency[7], in comparison to that on the plateau. Site environmental data has been collected over a span of 10 years. Based on this data, and comparisons with other telescopes in the area, survival and operating conditions were set for the project (see Table 2). The typical cold air temperature negates the need for air-conditioning systems for cooling, simplifying the design. The location, on a shoulder just east of the summit, partially shelters the site from the predominately westerly winds (see Figure 2).

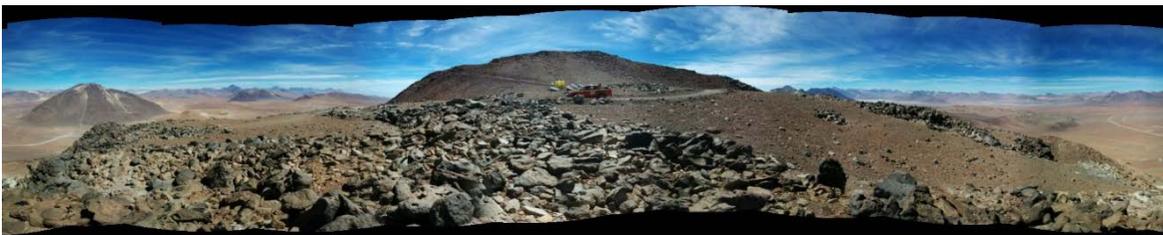

Figure 2. Panoramic view of the CCAT high site, looking from the eastern corner. West is roughly at the center of the picture. Cerro El-Chascón can be seen on the left and the Chajnantor plateau (location of ALMA) is below.

Table 2. Operating and survival environmental conditions of the telescope for some key parameters. The lower number for operating wind speed is for no loss in performance.

| Telescope Environmental Conditions | | |
| --- | --- | --- |
| Parameter | Operating | Survival |
| Air pressure | 50 to 53 kPa | 50 to 105 kPa |
| Wind speed | < 9 m/s, < 15 m/s | < 69 m/s |
| Air temperature | -21 to +9°C | -30 to +25°C |

## 2.2 Optics

The optics are unique for this size (~6 m) and type of telescope (crossed-Dragone). The optical design is based on one put forward by Niemack[2] and described in detail in Parshley et al.[4] The basic telescope optics are presented here (Figure 3). The system consists of two reflectors, arranged in a crossed-Dragone configuration providing an unobstructed aperture and low cross polarization. The primary and secondary mirrors are defined by high order polynomials instead of simple conics in order to correct for coma. The optics deliver a wide FoV and a fairly flat focal plane. The relative surface area of the primary to the secondary is 1.12, and this significantly drives the design of the telescope and its novel arrangement. Other submm telescope mirrors typically have area ratios anywhere from roughly 10 to 100.

The optics are oriented such that the chief ray from the secondary mirror (M2) is collinear with the elevation axis. This minimizes M2 performance changes with changing elevation angle, since the force of gravity acts mostly in-plane, regardless of the telescope's orientation. The design also allows for a 0-180° elevation (EL) range, which is useful for checking systematics by "flipping" the beam on the sky. There is a single Nasmyth position for large instruments (< 6 metric tons) and two bent Nasmyth positions (using fold mirrors) for small instruments (< 1.5 metric tons). The instruments only move with the azimuth structure and do not tip in elevation, thus simplifying instrument design, improving stability, and allowing for a simple optional third axis of motion, an instrument rotator, to help with systematics.

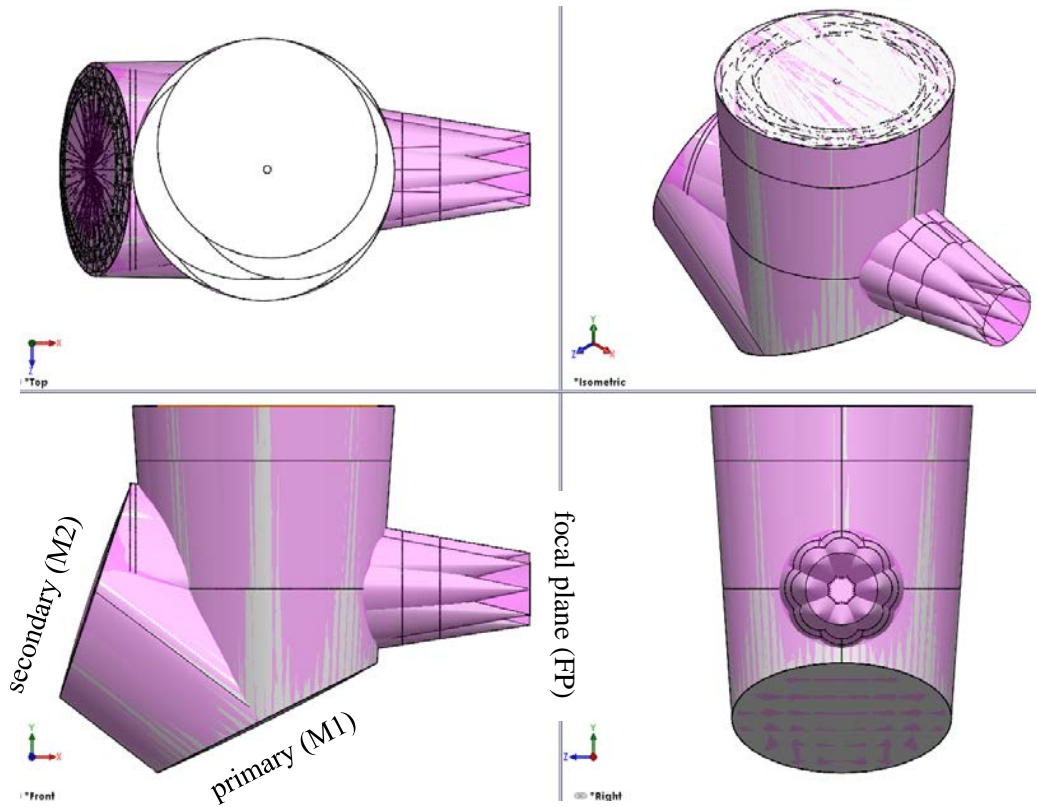

Figure 3. Beam model for the telescope. Rays (grey) for eight field positions at the edges of the field of view are overlaid and a composite beam keep-out model (magenta) is produced for the mechanical design. A 100 mm minimum beam clearance is added to the beam keep-out model to ensure adequate clearance to any structural elements.

## 2.3 Layout

Figure 4 shows the internal layout of the telescope. The telescope must provide a rigid support structure for the optics and focal plane instrumentation, as well as provide adequate space for support equipment. For CCAT-prime, three focal plane instrument positions are allowed. Instrument Space 1 is the "straight" Nasmyth position, for large FoV instruments, located on the third floor, and has direct access to the telescope focal plane. Instrument Spaces 2A and 2B, which are in the bent Nasmyth positions, are for smaller instruments, and access the telescope focal plane via fold mirrors and/or reimaging optics that send the beam down to the second floor. The Process Space will house compressors for cryocoolers and the Electronics Space provides room for control equipment, backends, and detector readout electronics. Having all science equipment located inside the yoke eliminates cable wraps between spaces. Not having the focal plane instruments move with elevation simplifies cryostat design. An overhead crane services Instrument Space 1, and smaller hoists are available on the 1$^{st}$ and 2$^{nd}$ floors for transfering equipment.

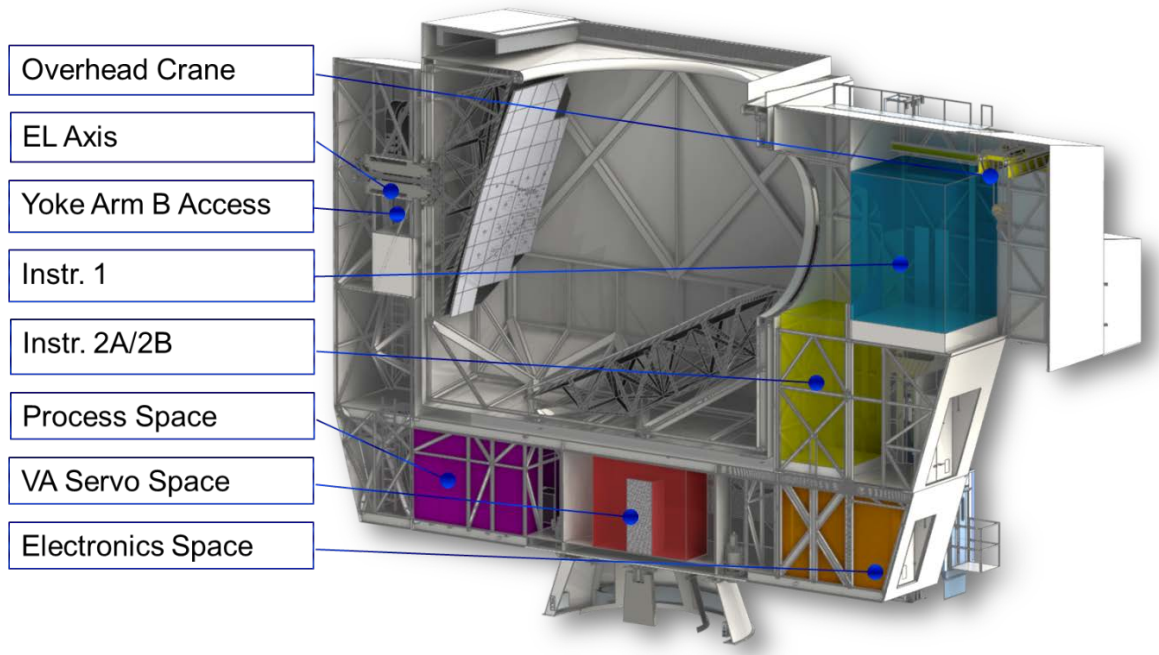

Figure 4. Cross section through the CCAT-prime telescope highlighting science equipment spaces and access.

## 2.4 Thermal management

An essential requirement of the telescope structure is to maintain the alignment, both form and position, of the optics across the range of operating conditions. It was decided to try to meet these requirements with a passive rather than active (i.e., mirror panel actuators) system, insensitive to temperature changes and gradients. Although a passive design solution for the telescope can mitigate effects for uniform temperature changes, temperature gradients in the structure can still adversely affect the performance. As such, a ventilation system (Figure 5) is being designed to minimize gradients in the structure. Both the elevation housing and the yoke will be cladded with insulating panels. The yoke is further divided into three pieces; two arms and a traverse. The arms and the elevation housing will be vented with ambient air in order to maintain a much more uniform temperature. The traverse will be temperature regulated to operate between 10 to 20° C by circulating air within the traverse for uniformity and adding outside air for cooling. Heaters will be installed but are expected to see limited use under normal operations as waste heat from compressors and electronics will help warm the space.

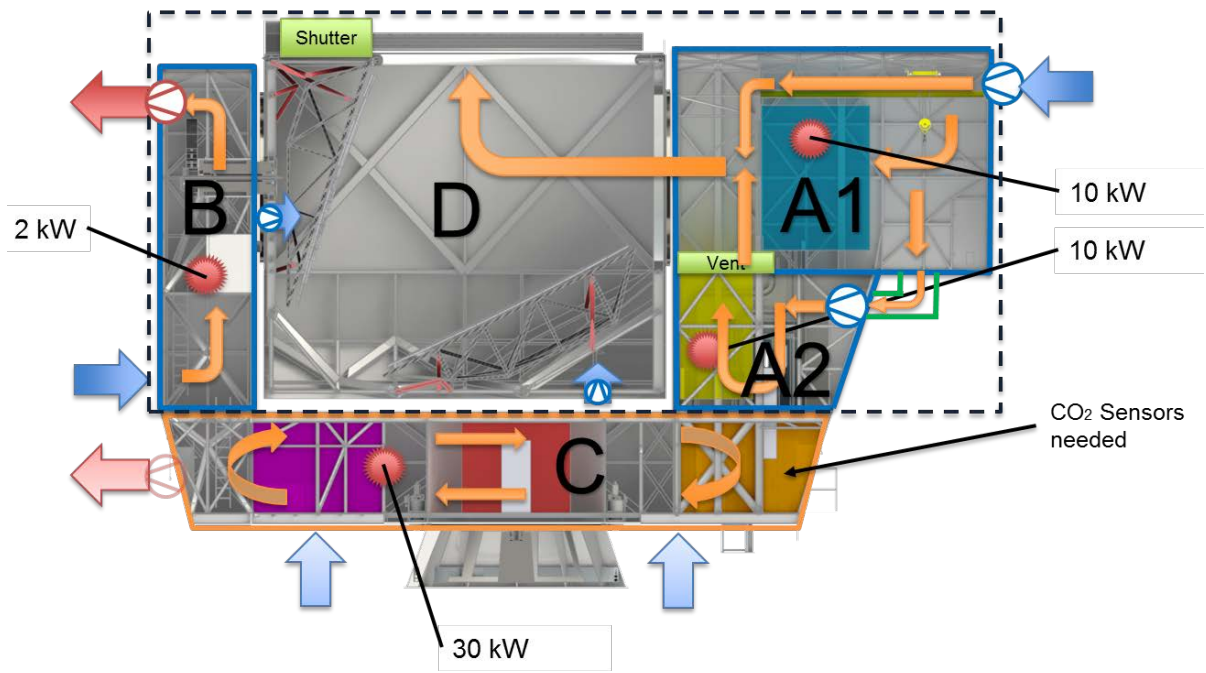

Figure 5. Ventilation concept for CCAT-prime. The yoke traverse is insulated and maintained between 10 to 20° C while the yoke arms and elevation housing are ventilated with the goal of improving temperature uniformity under varying environmental conditions. Blue arrows are fresh air intakes and red are warm exhaust. Heat sources and nominal flow paths are highlighted. Waste heat sources include 2 kW for the elevation drives, 10 kW for focal plane instruments, and 30 kW for all equipment in the traverse (e.g., compressors, backends, drive cabinets, etc.).

## 3    SUBASSEMBLIES AND COMPONENTS

This section details the major components critical to the mechanical structure, starting at the reflecting surfaces and ending at the ground. Each component is optimized individually before being combined into the full telescope model for further total optimization. This technique allows for parallel path development and leads to a good understanding of the sensitivity of the system to design changes.

### 3.1    Mirror panels

The panels (Figure 6) form the reflecting surfaces of the telescope optics, the position and tolerance of which must be carefully managed to ensure performance meets or exceeds requirements. The panels are light-weighted aluminum plates machined from a solid block. All panels have a common backside structure consisting of ribs and adjuster connection points. This simplifies the design and manufacturing. The front-side reflecting surface is unique for each panel due to the coma-corrected optics which lack the circular symmetry of a typical off-axis design. There are 87 panels on the primary reflector (M1) and 78 panels on the secondary reflector (M2). A single panel covers about half a square meter and weighs roughly 5 kg. Each panel is located on the BUS with eight adjusters; three in-plane and five vertical. Low order panel distortions from manufacturing can be partially compensated for with the vertical adjusters because they over-constrain the panel.

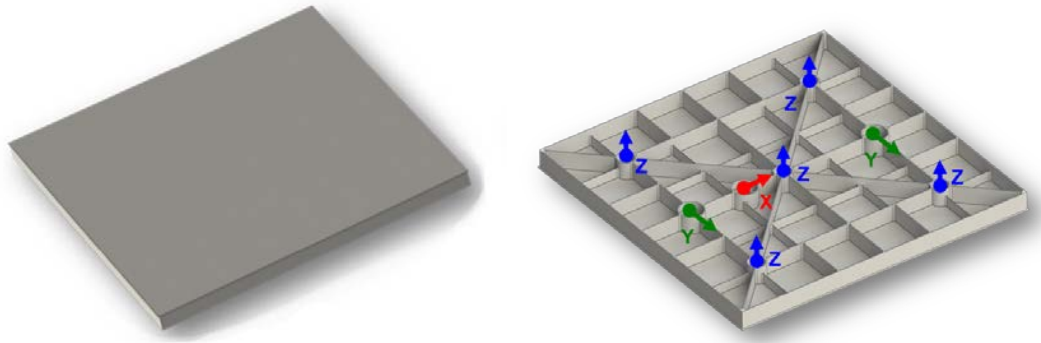

Figure 6. Mirror panel views, reflecting surface (left), backside (right). The panel is ~700 mm on a side and significantly light-weighted from a solid block of aluminum. Eight adjusters locate the panel. The five z axis adjusters allow for some compensation of low order distortions.

### 3.2 Backup structure (BUS)

The mirror backup structure (BUS) supports the panels in a stable configuration under all operating loading conditions; wind, thermal and gravity. Figure 7 shows the BUSes for both M1 and M2. They utilize a CFRP-AL-CFRP sandwich top plate, providing local stiffness for panel adjusters. CFRP tubes are bonded through the top plate to follow the mirror curvature. The top plate is in turn supported by a CFRP truss, which brings the load down to discrete interface points for isostatic mounting inside the elevation housing. The isostatic mount connections will have a tuned CTE to minimize the HWFE and PE resulting from rigid body shift and rotations of the optical elements due to changing thermal conditions.

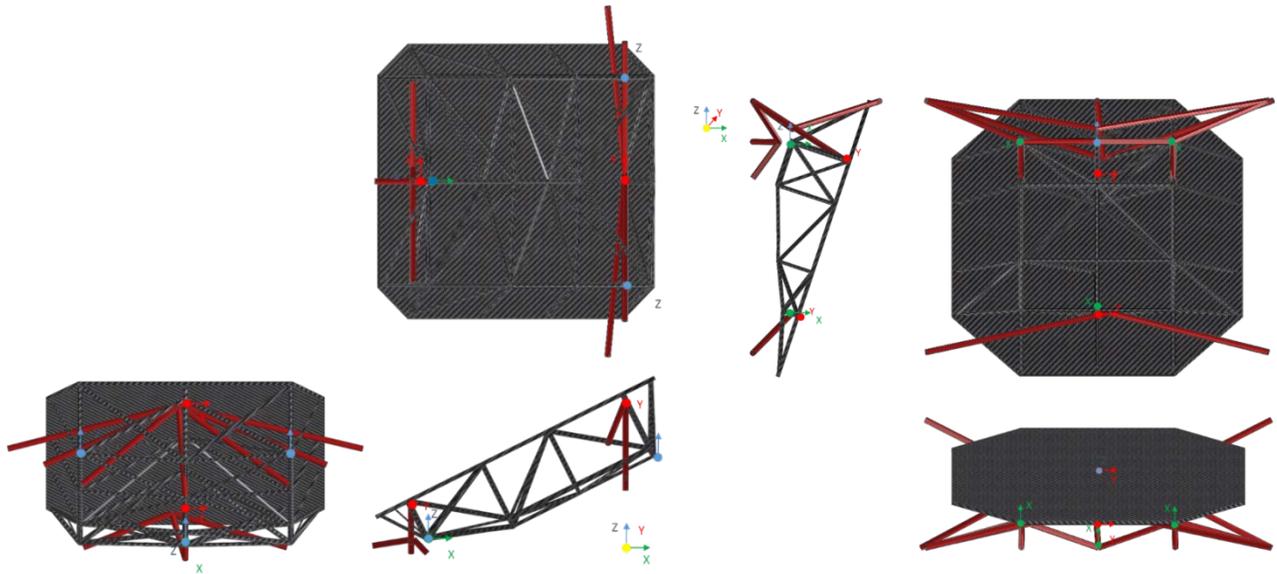

Figure 7. M1 (left) and M2 (right) CFRP BUS designs (panel adjuster tubes are not shown). The red elements are tuned CTE beams that minimize the HWFE and PE due to imperfect mirror alignment from thermal effects.

### 3.3 Elevation housing

The elevation housing (Figure 8) contains M1 and M2 inside an Invar framework with insulated exterior panels and a retractable shutter for environmental protection. In order to balance the structure about the EL axis, counter weights are added to the top to offset the effect of M1 on the center of mass. The SO LAT performance requirements are more relaxed than the CCAT-prime telescope allowing its elevation housing to be constructed from steel.

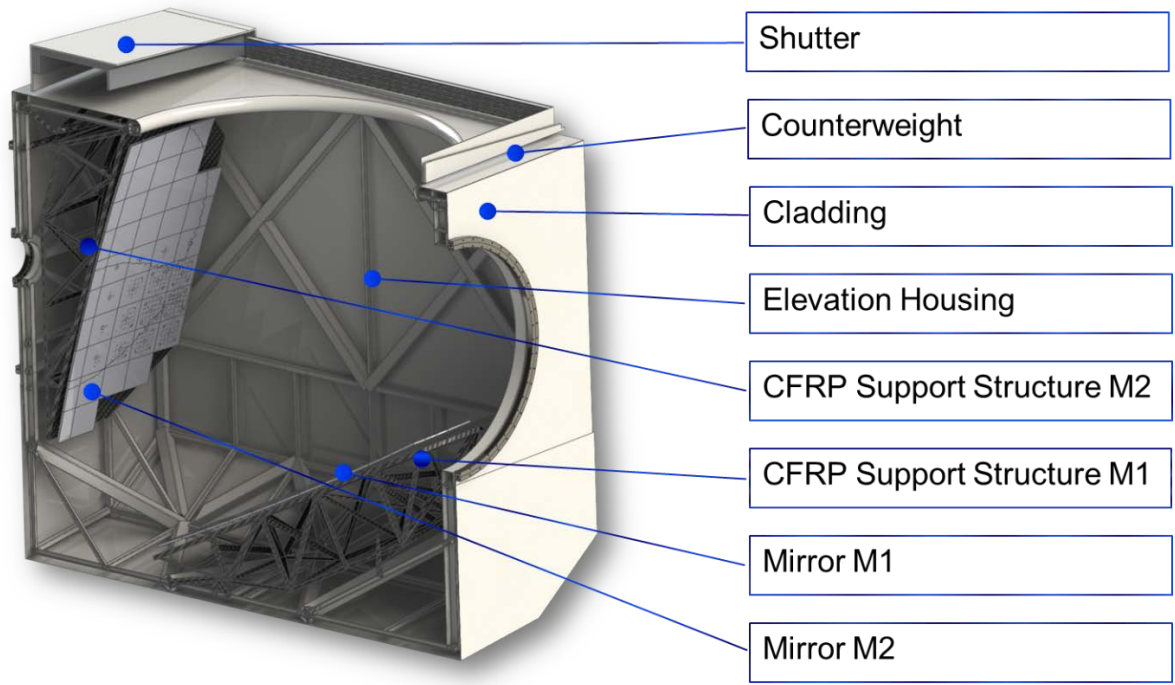

Figure 8. Elevation housing in cross-section through the telescope mid-plane with mirrors installed.

### 3.4 Yoke mount

The yoke (Figure 10) supports the elevation housing on its bearings, allowing 180° of travel. The yoke in turn rotates in azimuth on the support cone up to 540°. The yoke is an insulated steel framework. It contains space for focal plane instruments, backend electronics, compressors, and telescope control equipment. A lift is located on one side for moving personnel between floors to minimize strain on people working at the high altitude site. The elevation encoder and drives are located opposite the focal plane.

### 3.5 Support cone

The support cone is fixed in azimuth and rigidly connects the entire telescope through the azimuth bearing to the foundation. It also houses the azimuth cable wrap, limit switches and encoders. It is a steel structure, bolted to the foundation via an anchor ring secured by cast-in-place concrete. See Figure 9.

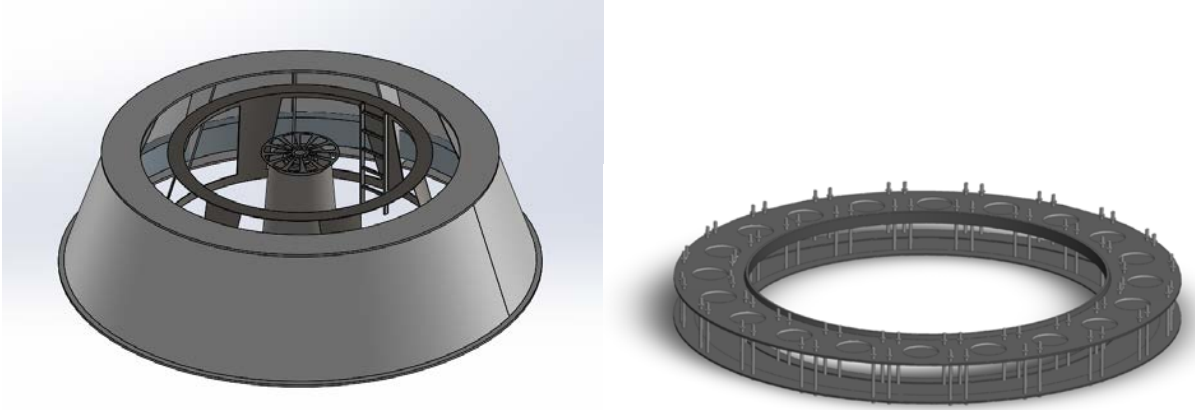

Figure 9. Support cone (left) and anchor ring (right). The anchor ring will be cast in place in the concrete foundation.

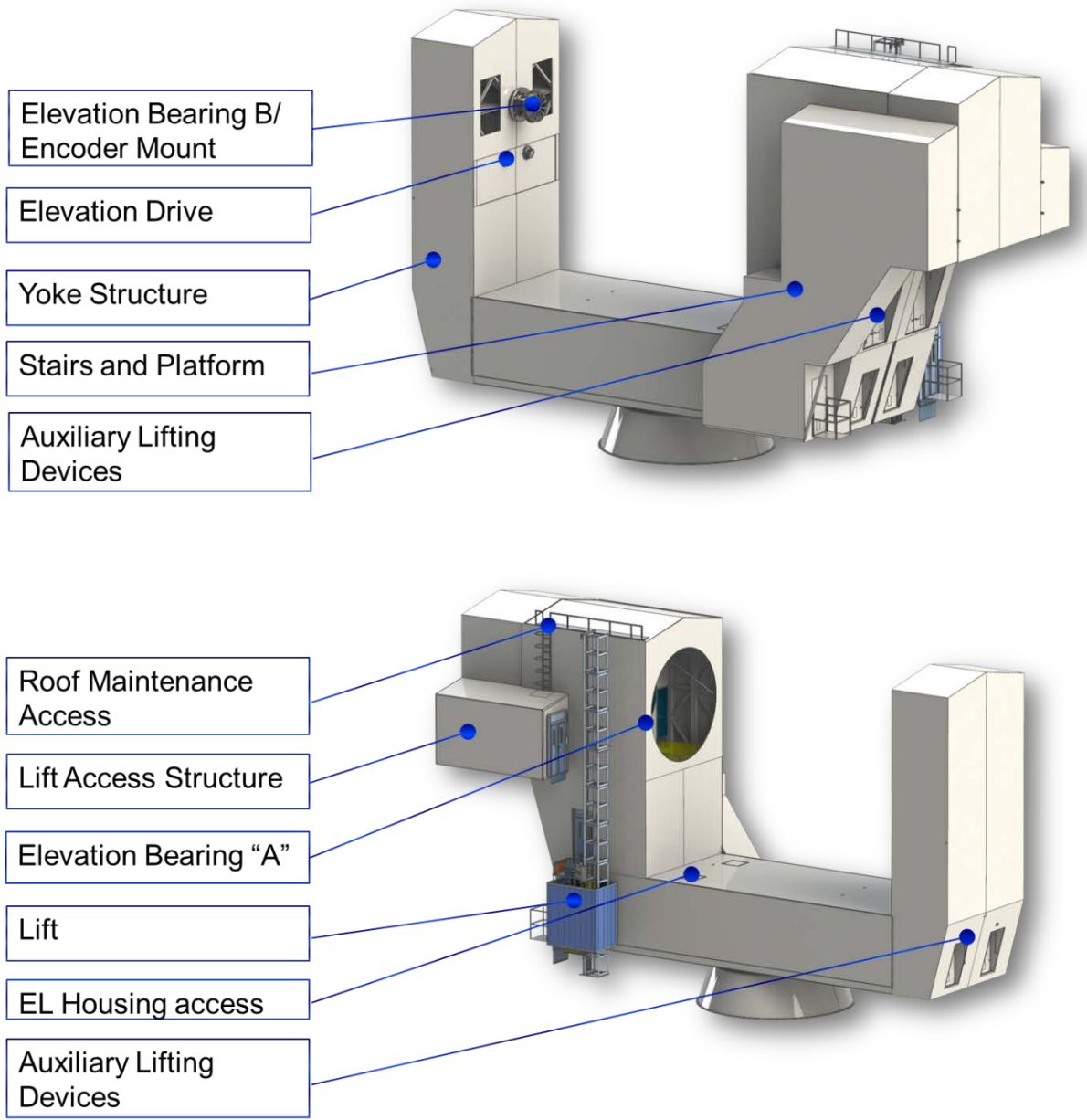

Figure 10. Two views of the yoke structure.

## 4  SIMULATED PERFORMANCE

In order to assess the quality of the design and allow for optimization, the overall performance of the telescope is simulated under various load cases. The steady-state effects of gravity (Figure 11), thermal gradients (Figure 12), uniform temperature changes, and wind are applied to the model. Relative deformations are calculated from the results and converted into HWFE (Table 3) and PE (Table 4). The telescope emissivity and dynamic performance are also calculated.

To ensure physically meaningful load cases, the wind loads are computed by an extensive computational fluid dynamics / fluid-structure interaction analysis study (Figure 13 and Figure 14) and the temperature loads (Figure 15) are, inter alia, computed by transient analysis. The outcome of these studies are the resulting wind pressure distribution on the telescope and the temperature distribution of the resulting thermal gradients on the structure.

The simulation model contains numerous individually positioned mass points for the realistic consideration of the masses of the added science equipment (Figure 16).

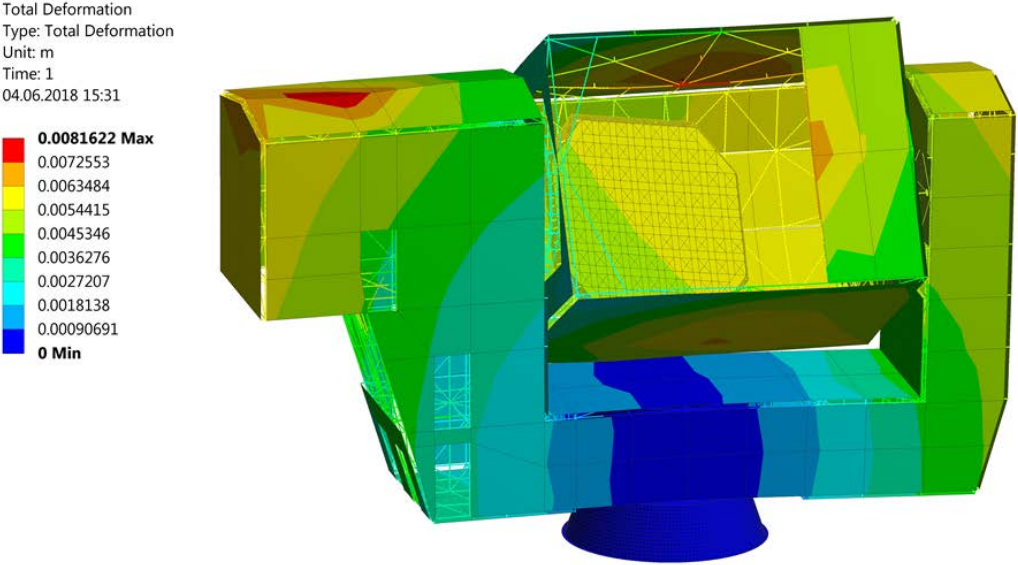

Figure 11. Deformed shape of the telescope structure due to gravity at 30° EL.

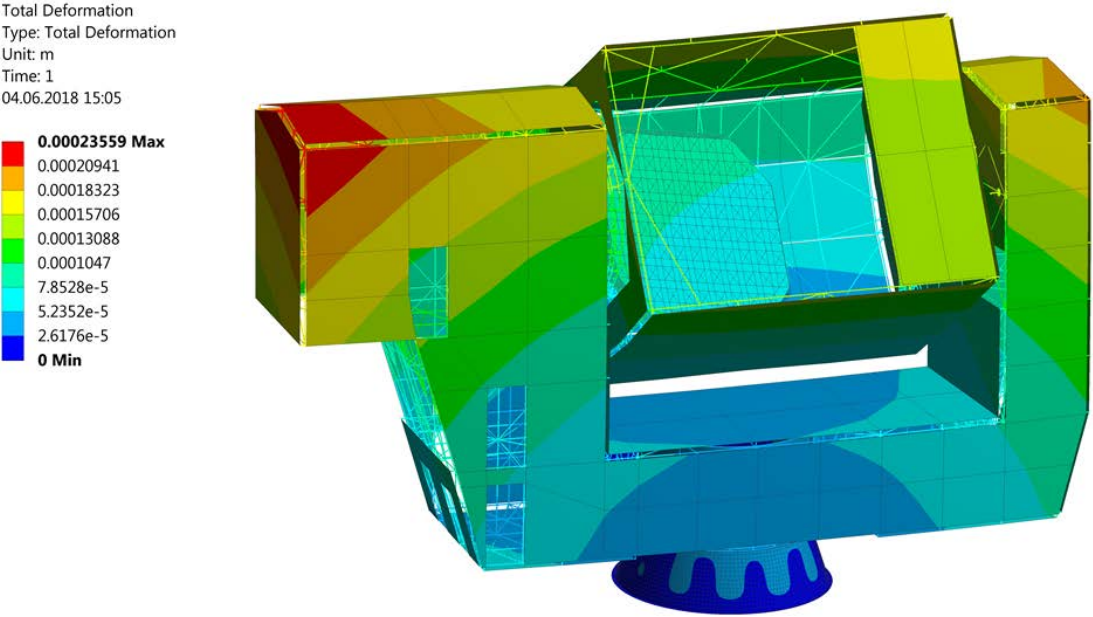

Figure 12. Deformed shape of the telescope structure due to a thermal gradient front-to-back at 30° EL.

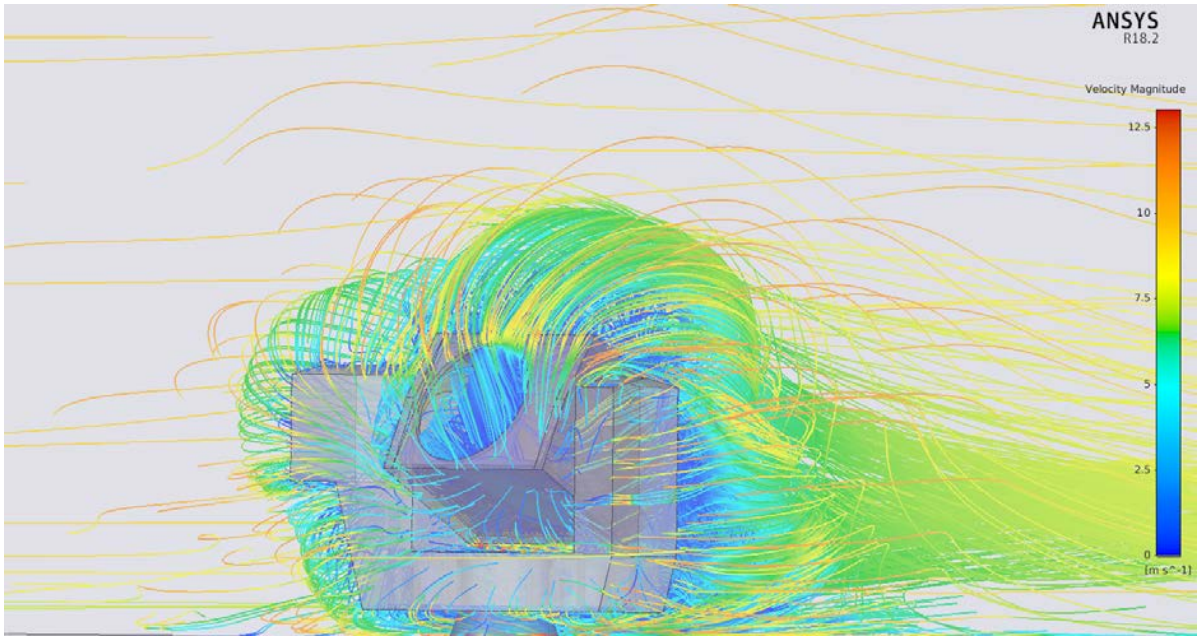

Figure 13.  Wind velocity magnitude visualized by a streamline plot, for initial wind speed of 9 m/s (the operating baseline). The telescope is oriented -30° AZ, 30° EL to the wind direction.  An air density of 0.646 kg/m$^3$ is used (average conditions).

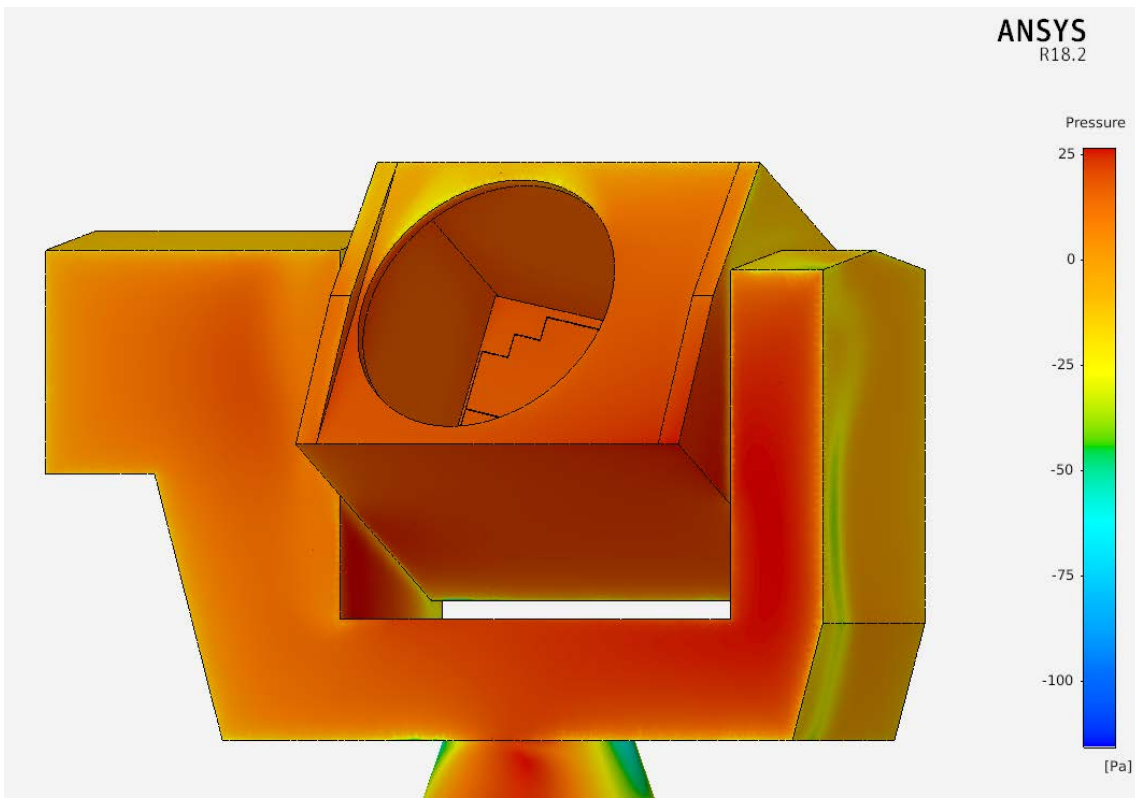

Figure 14.  Resulting wind pressure distribution for a wind speed of 9 m/s, same telescope orientation and wind blow direction as Figure 13.

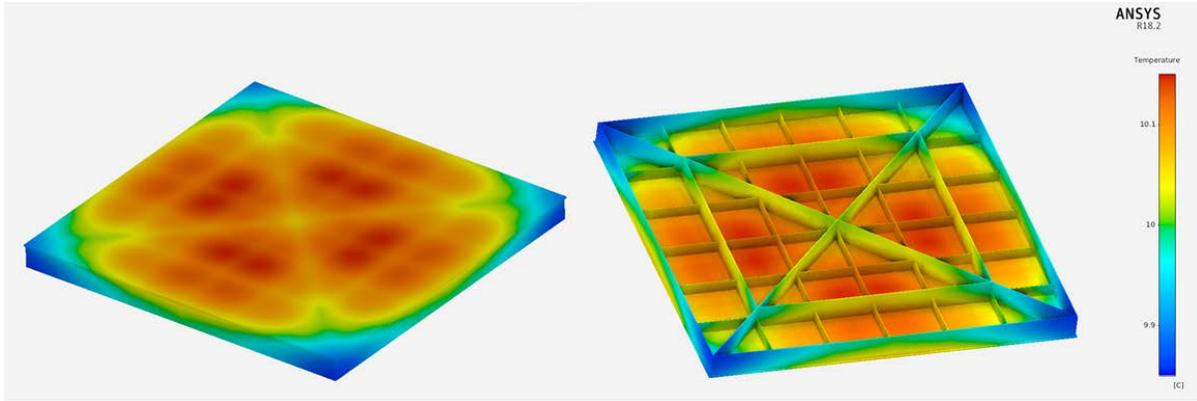

Figure 15. Thermal gradient on a single mirror panel computed by a transient analysis after 300 s. A heat load is applied to the mirror surface and convective cooling applies to all surfaces. The peak-to-peak value is approximately 0.3° C.

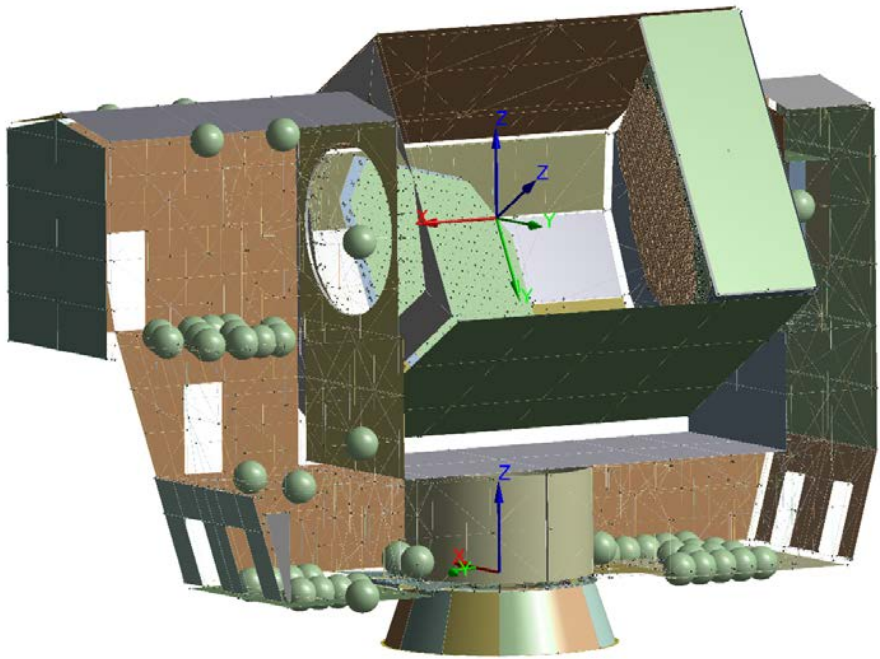

Figure 16. View inside of the of the entire telescope structure finite element model (30° EL) showing individual point masses (spheres) for telescope (e.g., drives, cabinets) and science equipment (e.g., instruments, compressors, electronics racks).

### 4.1 Half wavefront error

Table 3 shows the system HWFE performance by major component and also by error source. It shows that gravity is the single largest contributor at roughly 7 μm. As a comparison, the natural (von Hoerner) limits[8] for a 6 meter telescope under gravitation loading are roughly 3 and 11 μm for a CFRP and steel structure, respectively. Thermal effects, as well as alignment and manufacturing tolerances also contribute significant to the overall HWFE. The effect of wind seems to be almost negligible. Presently, the design successfully achieves the HWFE requirement. As the design matures, this performance will be continually checked. It is hoped we will do better than the 11 μm rms requirement and come close to our ultimate goal of 7 μm rms.

Table 3. HWFE by component and error type. Note that zero under alignment (Align.) and manufacturing (Mfg.) means the error is (or can be made) a negligible amount. All errors are in μm rms.

| Error Type \ Component | Gravity | Temp, grad | Temp, soak | Wind | Align. | Mfg. | Margin | Component Total |
|---|---|---|---|---|---|---|---|---|
| Mirror config. | 4.2 | 0.3 | 3.5 | 0.0 | 2.0 | 0.0 | 1.0 | 5.9 |
| M2 BUS | 2.3 | 1.0 | 0.1 | 0.2 | 2.0 | 0.0 | 1.0 | 3.4 |
| M1 BUS | 5.1 | 1.0 | 0.1 | 0.4 | 2.0 | 0.0 | 1.0 | 5.7 |
| M2 panel | 0.8 | 2.2 | 0.9 | 0.2 | 0.0 | 3.0 | 1.0 | 4.0 |
| M1 panel | 0.8 | 2.2 | 0.9 | 0.2 | 0.0 | 3.0 | 1.0 | 4.0 |
| Error Type Total | 7.1 | 3.4 | 3.7 | 0.5 | 3.5 | 4.2 | 2.2 | Telescope Total: **10.5** |

### 4.2 Pointing error

The overall pointing error is the vector sum of the elevation and cross-elevation errors, which in turn can be broken down into repeatable and random errors. Errors from gravitation deflection are highly repeatable and can be compensated for with a pointing model. Errors from transient effects such as wind and thermal changes are difficult to compensate and must be minimized via careful mechanical design. Only the pointing stability performance is out of specification at this time. We believe further optimization and compensation efforts will correct this. Table 4 shows the pointing error performance.

Table 4. Pointing errors, all errors are in arcsec rms.

|  | 9 m/s requirement | 9 m/s performance | 15 m/s performance |
|---|---|---|---|
| Blind pointing | 6.9 | 5.4 | 5.8 |
| Offset pointing | 2.7 | 1.6 | 2.7 |
| Scan pointing, sector 1 | 1.4 | 1.4 | 2.5 |
| Scan pointing, sector 2 | 1.9 | 1.5 | 2.6 |
| Scan following | 6.9 | 1.6 | 1.6 |
| Pointing stability | 1.4 | 1.7 | 1.9 |

### 4.3 Emissivity

The telescope blockage is calculated from the mechanical model. Since the optics are offset and unobscured, telescope blockage comes solely from the gap between panels. For the worst case operating condition, T = -21° C, the gap is calculated to be approximately 1.5 mm. Combined with the reflectivity of aluminum as a function of wavelength, the total emissivity of the telescope is shown in Figure 17.

### 4.4 Modal Analysis and Servo Simulation

A locked rotor analysis (Figure 18) gives information about the stiffness of the overall telescope structure including the drive system. Furthermore, for the simulation of the telescope dynamic behavior, a free rotor analysis is executed in order to generate the mechanical part of the dynamic simulation model. The model generation is performed by a modal analysis technique.

The purpose of simulating the dynamic performance of the telescope helps inform the design process regarding impacts to the dynamical behavior, assess tracking and scanning accuracy, determines disturbance (e.g., wind) behavior of the telescope, and ultimately can reduce commissioning time on site by pre-optimization of controller parameters.

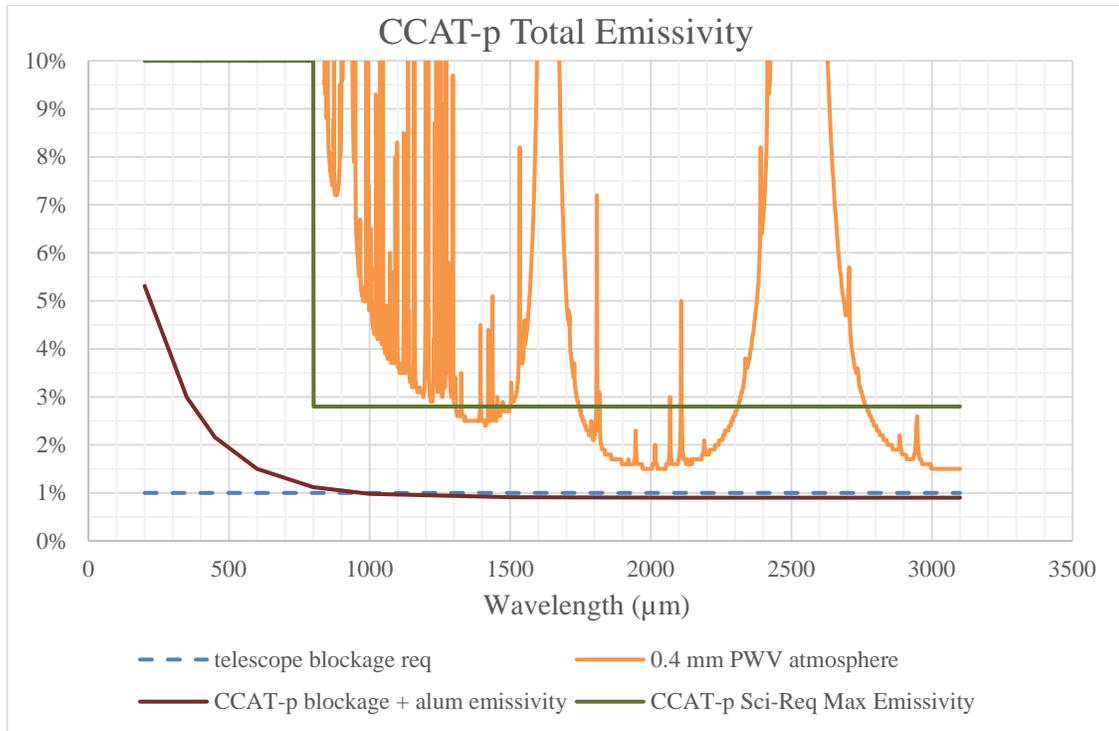

Figure 17. Plot of telescope blockage and science emissivity requirement, calculated telescope performance at -21° C, and comparison with the sky emissivity of the Chajnantor site under average conditions for the summit.

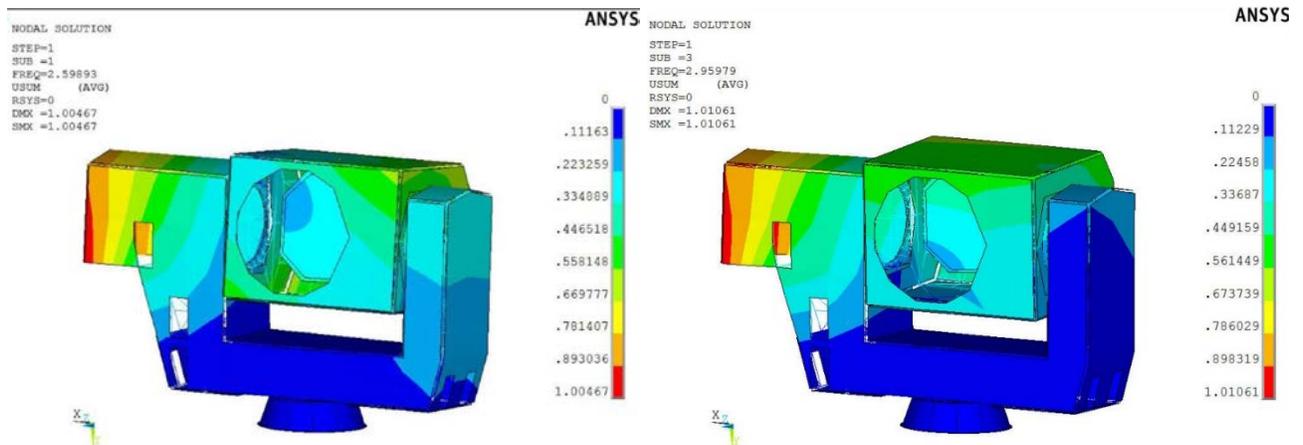

Figure 18. Left – 2.6 Hz locked rotor azimuth (1st mode) for 0° elevation. Right – 3.0 Hz locked rotor elevation (3rd mode) for 0° elevation. Modes are not pure in the sense that the axes motions are somewhat coupled. Stiffening the yoke structure should improve this.

## 5    CONCLUSION

We have completed the preliminary detailed design and simulation for a 6 meter crossed-Dragone telescope with a primary to secondary area ratio near unity. Although the site requirements and science goals are challenging, this design is capable of meeting or exceeding the performance requirements for HWFE, PE, emissivity and dynamics. A clear advantage of having the optics "buried" inside the mount is lower wind loading on the mirrors as they are effectively inside an enclosure, and the typical secondary support structure is eliminated. It also facilitates spillover control, reducing ground pickup. The design has an integrated shutter that will provide protection from sun, rain, snow, and ice during poor observing conditions.

The chief ray from M2 in the crossed-Dragone optical layout is used as the elevation axis in an elevation-over-azimuth mount with a single Nasmyth position for large instruments and two bent Nasmyth positions (with fold mirrors) for small instruments. The instruments only move with the azimuth structure and do not tip in elevation, thus simplifying instrument design, improving stability, and allowing for a simple optional fourth axis of motion, an instrument rotator, to help with systematics. Due to the symmetrical nature of the mount, the bore sight can be "flipped" on sky by rotating in elevation beyond zenith (> 90° EL) and coming back around 180° in azimuth.

The Simons Observatory has adopted the same telescope design for their Large Aperture Telescope. The Mizuguchi-Dragone optical design and the large FoV makes it ideal for observations of the polarization of the cosmic microwave background radiation at millimetric wavelengths. While the design is practically identical, SO opted to use steel instead of Invar for their LAT as, operating at longer wavelengths, the pointing and HWFE requirements are less stringent. Both CCAT-prime and SO LAT are planning for first light in 2021.

**Acknowledgements:** CCAT-prime funding has been provided by Cornell University, the Fred M. Young Jr. Charitable Fund, the German Research Foundation (DFG) through grant number INST 216/733-1 FUGG, the Univ. of Cologne, the Univ. of Bonn, and the Canadian Atacama Telescope Consortium.